\documentclass[conference]{IEEEtran}
\usepackage{graphicx} 
\usepackage{todonotes}
\usepackage[utf8]{inputenc}
\usepackage{hyperref}
\usepackage{multirow}
\usepackage{tcolorbox}
\usepackage{amsmath}
\usepackage[table,xcdraw]{xcolor}
\usepackage{balance}
\usepackage{cite}

\usepackage[normalem]{ulem}
\useunder{\uline}{\ul}{}

\title{On Assessing the Relevance of Code Reviews Authored by Generative Models}

\author{
	\IEEEauthorblockN{Robert Heumüller}
		\IEEEauthorblockA{
				Otto von Guericke University Magdeburg, Germany\\
				robert.heumueller@ovgu.de}
		\and
		\IEEEauthorblockN{Frank Ortmeier}
		\IEEEauthorblockA{
				Otto von Guericke University Magdeburg, Germany\\
				frank.ortmeier@ovgu.de}
	}

\begin{document}

\maketitle

\begin{abstract}
The use of large language models like ChatGPT in code review offers promising efficiency gains but also raises concerns about correctness and safety.
Existing evaluation methods for code review generation either rely on automatic comparisons to a single ground truth, which fails to capture the variability of human perspectives, or on subjective assessments of "usefulness," a highly ambiguous concept.
We propose a novel evaluation approach based on what we call multi-subjective ranking.
Using a dataset of 280 self-contained code review requests and corresponding comments from CodeReview StackExchange, multiple human judges ranked the quality of ChatGPT-generated comments alongside the top human responses from the platform.
Results show that ChatGPT’s comments were ranked significantly better than human ones, even surpassing StackExchange's accepted answers.
Going further, our proposed method motivates and enables more meaningful assessments of generative AI's performance in code review, while also raising awareness of potential risks of unchecked integration into review processes.
\end{abstract}
\section{Introduction}
Code review automation represents a critical area of research, given the substantial time and effort required for review processes in practice.
The efficiency gains recently made possible by generative large language models (LLMs) like OpenAI's ChatGPT or GitHub Copilot are quickly changing software development practice \cite{Noy2023, Peng2023}.
Yet, the unchecked application of LLMs in software development also presents many risks, including potential correctness and security concerns, so research into this topic is essential to enable safe usage.

In terms of LLMs in code review automation, research has applied ChatGPT, ChatGLM3 and LLAMA to different tasks such as review necessity prediction, review comment generation and code refinement\cite{Tufano2024, Lu2023, Li2025}.
So far, code review automation approaches primarily evaluated the performance of generative models by either comparing their output to a ground truth using metrics like ExactMatches or BLEU \cite{Tufano20222291, Li20221035}, or by qualitatively assessing the usefulness of generated comments through human judges \cite{Gupta2018, Guo2019}.
However, these methods have important limitations.
The inherent variability, ambiguity, and occasional errors in human code review comments make it difficult to assert the exclusive correctness of any single "ground truth" response. 
Just as important, there are often many valid ways to express the same underlying semantic idea in a code review. 
Evaluation methods that rely heavily on lexical similarity metrics like ExactMatches or BLEU fail to account for this and are thus likely to fall short.
Furthermore, unlike human reviewers, who are typically developers familiar with the project, code review models so far have limited or no access to the broader context surrounding the change under review\cite{Zhou2023, Shuvo2023, Lu2023, Lu2023, Tufano20222291, Li20221035}. 
As a result, expecting them to reach the same conclusions as a human reviewer may be unrealistic.
\emph{Consequently, previous methodologies are not optimally suited to assess the relevance of  model-generated review comments against a human standard.}

{
\setlength{\abovecaptionskip}{2pt}
\setlength{\belowcaptionskip}{2pt}
\begin{figure}
	\includegraphics[width=\linewidth]{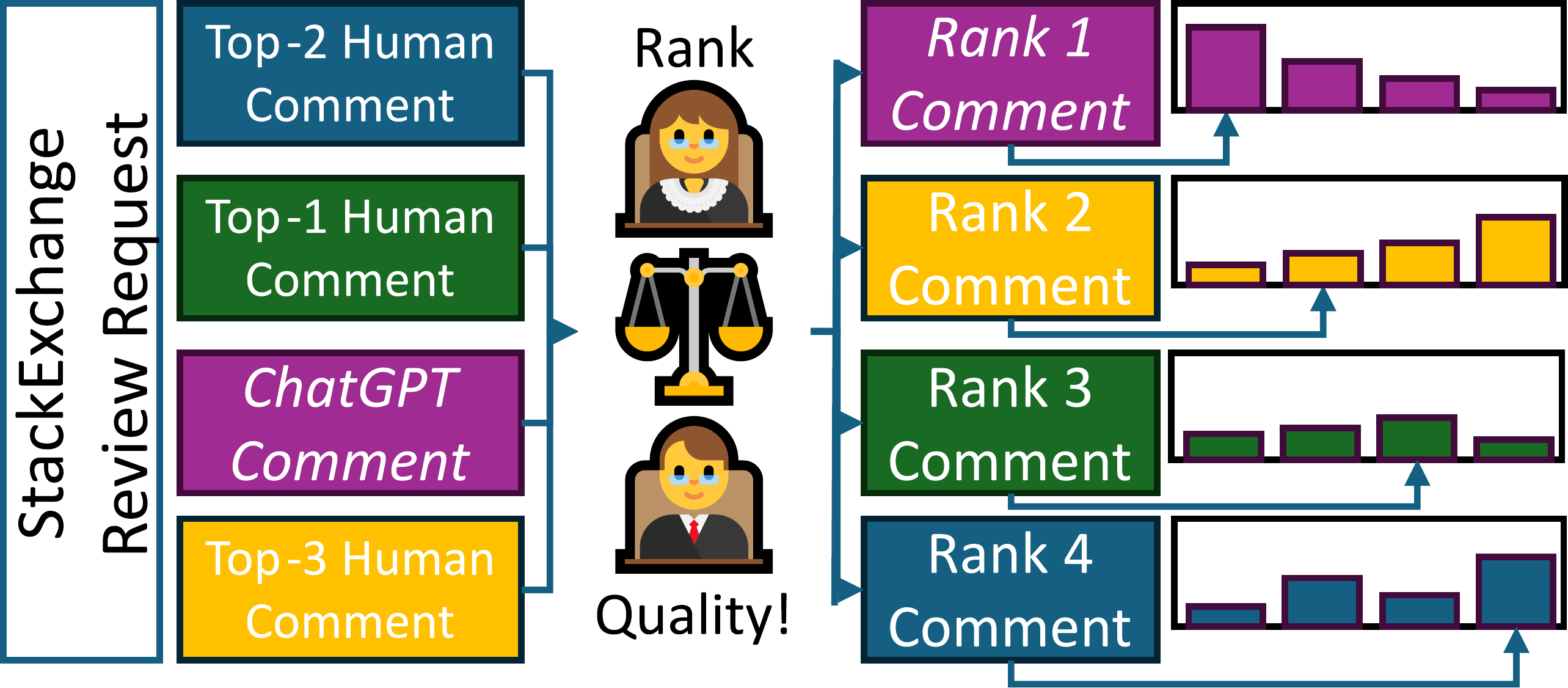}
	\caption{Multi-Subjective Ranking Analysis with 4 Judges\label{fig-overview}}
\end{figure}
}

In this paper, we present a method to overcome these challenges and report initial results from its application to ChatGPT 3.5 Turbo\cite{chatgpt-3.5-turbo}.
To this end, we conduct an empirical study on a dataset of 280 code reviews and their corresponding top-voted comments, derived from the CodeReview Stack Exchange platform.
In what we call a \emph{multi-subjective ranking experiment}, illustrated in Figure~\ref{fig-overview}, four human judges rank a ChatGPT-generated comment alongside \emph{the top three human comments} in the context of a \emph{self-contained review request requiring no additional context}.
The process of multi-subjective ranking, popularized especially by platforms like Stack Overflow, is widely accepted as a way of comparatively assessing the quality of multiple candidate solutions.

Our research hypothesis is thus: \emph{ChatGPT's reviewing capability can be considered relevant with respect to a human standard, if it consistently displaces one of the top three human review comments for requests on CodeReview Stack Exchange.}

To facilitate this study, we developed a web application called \emph{Rankr}, designed to crowdsource ranking data.
Our initial findings reveal two key insights.
\emph{Firstly, when excluding the generated comment, human rankers consistently and significantly agree with the platform's top-voted and accepted answers as the best of the human responses.}
This speaks to the validity of the rankings by our four judges.
\emph{Secondly, when including the generated comment, rankers significantly prefer the ChatGPT-generated comment over all human ones, statistically even surpassing the platform's accepted answer.}

Although we took measures to control for the style and tonality of generated reviews and randomized the order of presentation, there is a potential threat to construct validity, as it is sometimes possible to detect the generated review comment based on superficial features: ChatGPT's grammar and wording are notably more polished than typical human review comments.
Consequently, the high rankings of generated comments might be biased by a preference for such superficial features.
\emph{However, even if true, this actually highlights the dangers of unchecked use of generative models like ChatGPT for code reviews, underscoring the need for further research.}

Finally, these limitations apply only to initial results and we identified strategies to mitigate the possible confounding factors in future work.
We believe that \emph{the general method we propose, can be an important step towards gaining a deeper understanding of generative models' performance in code review and thereby help to ensure their safe usage.}
Our human-annotated dataset and evaluation code are available for replication and independent scrutiny\footnote{https://github.com/robert-heumueller-ovgu/repl-generative-review-relevance}.

\textbf{Related Work:} Generative large language models have previously been applied to code review comment generation.
This work demonstrated that both general-purpose LLMs (ChatGPT), fine-tuned LLMs (LLAMA-based) and comprehensive frameworks built around LLMs (ChatGLM) can at least match and even surpass the performance of models trained specifically for code review tasks\cite{Tufano2024, Lu2023, Li2025}.
While they also provide further insights, such as taxonomies of the types of code changes where models tend to succeed or fail, their evaluations rely primarily on simple lexical metrics—such as BLEU scores and ExactMatches and are limited to a single ground-truth comment per instance.
To the best of our knowledge, this is the first work to apply multi-subjective ranking with multiple human reference comments, offering much broader perspective on the perceived relevance and quality of model-generated comments.

\section{Methodology}
\textbf{Dataset:}
All experiments in this work are based on a dataset derived from StackExchange's the Code Review platform\footnote{\url{https://codereview.stackexchange.com/}}.
There, similar to the way people can ask questions, rate, and receive answers to software development related questions on StackOverflow, authors can submit self-contained code review requests.
Typically, these consist of a description of context and goals, a code snippet and one or multiple questions or specific aspects reviewers should comment on.
Reviewers can then answer to such requests by writing code review.
Then, all users can up- or down-vote reviews, i.e. perform multi-subjective ranking.
Finally, the original author can mark one review comment as their \emph{accepted answer}.
Note that this is not necessarily the top-voted answer, in fact, this is only the case for 50\% of the instances.

We extracted a total of 69,994 question–answer pairs and converted them into a machine-readable format. 
Each entry includes the original code review request and the top-three review comments by votes, including respective number of votes, and, if present, a label for the accepted answer.

Next, we had ChatGPT-3.5 turbo via its API generate a candidate comment for each review request using minimal prompt engineering ("Give me a useful code review to following question in a maximum of 200 words. Do not enumerate.")
We used ChatGPT-3.5 turbo, since at that time in 2024 it was the model with the best cost-performance-speed trade-off.

Finally, to keep the effort for manual ranking manageable, we filter instances based on the maximum length of 1000 characters for the review request, for each of the three human comments, and for the generated comment.
This way, we end up with a final dataset of 280 instances, each with a maximum length of 5000 characters.

\textbf{Ranking:}
To assess the quality of the generated code review comments, we conducted a human evaluation study involving four judges. 
These included both computer science students and professional developers, all with work experience in software development and code review processes. 
They were aged between 20 and 50. 
The ranking process was carried out using \texttt{Rankr}, a web-based tool we developed specifically for this type of evaluation \footnote{Rankr will be made available as open source software.}. 
For each of the 280 review instances, the judges were shown the complete review request along with the four anonymized and randomized candidate comments, and were asked to rank them from best to worst using a drag-and-drop interface.
Tied rank assignments were not allowed. 
The resulting dataset, included in our replication package, contains the ordinal ranks (1=best to 4=worst) assigned by all four judges for each comment. 
This structure is the basis for the statistical analysis in the following Section.
\section{Evaluation\label{sec-evaluation}}
\begin{figure*}[t!]
	\centering
	\caption{Histograms of Human Rank Distributions by Comment Type\label{fig-histograms}}
	\includegraphics[width=0.8\linewidth]{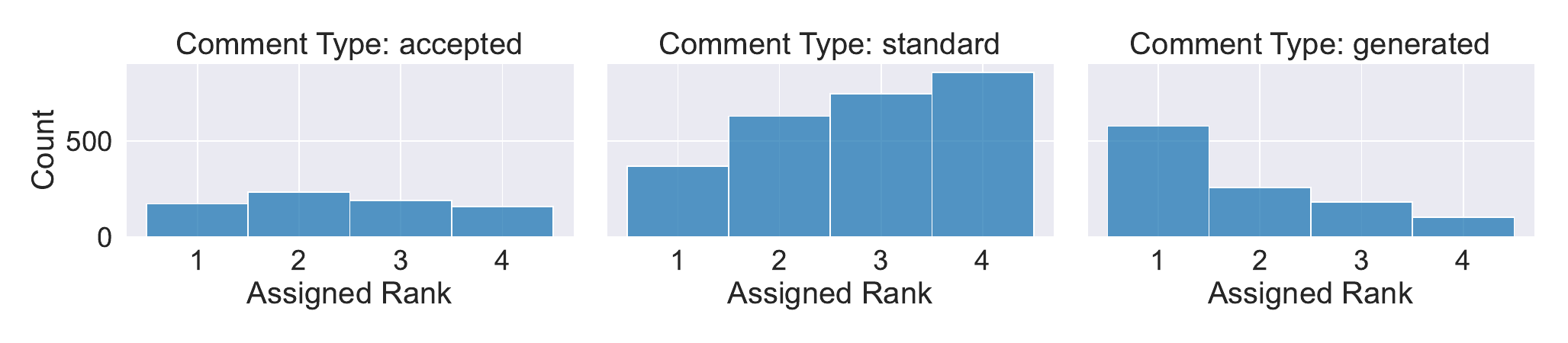}
\end{figure*}
\begin{table*}[h]
	\begin{minipage}{0.48\textwidth}
	\caption{Pairwise tests of our human-assigned ranks for different comment types. Significant p\_Values underlined. \label{table-significance-types}}
	\centering
	\begin{tabular}{l|lll}
		\textbf{Comment Type} & \textbf{Accepted} & \textbf{Standard} & \textbf{Generated} \\
		\hline
		\textbf{Accepted} & - & \underline{1.59e-16} & 1.00e+00 \\
		\textbf{Standard} & 1.00e+00 & - & 1.00e+00 \\
		\textbf{Generated} & \underline{1.55e-36} & \underline{2.25e-128} & - \\
	\end{tabular}
	\end{minipage}
	\begin{minipage}{0.48\textwidth}
	\caption{Pairwise tests of our human-assigned ranks for different StackExch. voting ranks. Significant p\_Values underlined.\label{table-significance-voting}}
	\centering
	\begin{tabular}{l|l|l|l|l}
		\textbf{Voting} & \textbf{Rank 1} & \textbf{Rank 2} & \textbf{Rank 3} \\
		\hline
		\textbf{Rank 1} & -                      & {\ul 1.73e-09}         & {\ul 1.42e-14}                    \\
		\textbf{Rank 2} & 1.00e+00               & -                      & 3.93e-02                             \\
		\textbf{Rank 3} & 1.00e+00               & 9.61e-01               & -                                  \\
	\end{tabular}
	\end{minipage}
\end{table*}

To begin our evaluation, we visualize in Figure~\ref{fig-histograms} the distribution of human-assigned ranks for each comment type.
The histograms reveal two key trends.
First, for StackExchange’s accepted answers (left histogram), the rank assignments are more uniformly distributed compared to those for standard human comments (middle histogram).
In the latter case, higher (i.e., worse) ranks are assigned with increasing frequency, suggesting that annotators consistently recognized the superior quality of the accepted answers.
Second, for the generated comments (right histogram), the rank assignments show an inverse pattern to the standard comments: the best rank was assigned most frequently, with decreasing frequencies for worse ranks.
In this way, the generated comments already appear to even exceed the accepted answers in perceived quality, at least based on rank distributions.
In the following we analyze these observations statistically in two experiments.

\vspace{0.25em}\noindent\textbf{Experiment 1: Alignment with StackExchange Judgments}\\
In the first experiment, we examine how well the rank annotations made by our judges align with StackExchange’s own indicators of comment quality. This includes two comparisons:

\begin{enumerate}
	\item We compare the ranks assigned to the \textit{accepted answer} versus those assigned to other human-written (standard) comments.
	\item We compare the ranks assigned to answers based on their voting rank on the platform (i.e., top-voted, second-voted, and third-voted).
\end{enumerate}

To determine whether our judges consistently rated accepted answers or higher-voted answers as better (i.e., assigned lower ranks), we performed pairwise, left-sided significance tests using the Mann–Whitney U test. This non-parametric test is well-suited to our setup, as it does not assume normality and is appropriate for ordinal rank data \cite{497e1044-d5b0-30a9-b230-3ca0f10d6f6c}. We used a standard significance level of \(\alpha = 0.01\)\cite{Tufano20222291}.

When performing multiple statistical tests, it is generally a good practice to apply a Bonferroni correction to avoid inflated Type I error rates\cite{dunn1961multiple}. 
However, in our case, all significant comparisons resulted in p-values smaller than \(10^{-9}\), which are orders of magnitude below the significance threshold. We can thus be confident that these results are not due to chance.

Table~\ref{table-significance-types} summarizes the results of the pairwise significance tests on judge-assigned ranks across different comment types: accepted, standard (non-accepted), and generated. In the first row, for instance, we test the null hypothesis that the ranks assigned to accepted answers are greater (i.e. worse) than or equal to those of standard or generated comments.

Table~\ref{table-significance-voting} shows the results of similar tests, this time based on StackExchange's voting ranks. In the first row, we test the null hypothesis the ranks assigned to top-voted comments are greater (i.e. worse) than or equal to second- and third-voted ones, and so on.

From Table~\ref{table-significance-types}, we observe that accepted comments are ranked significantly better than standard comments.

From Table~\ref{table-significance-voting}, we observe that top-voted comments are ranked significantly better than both second- and third-voted comments. However, the difference between second- and third-voted comments is not statistically significant. To investigate this, we examined the median number of votes per rank: top-voted comments had a median of 6 votes, second-voted had 3 votes, and third-voted had 2 votes. The difference in votes between ranks 1 and 2 is therefore twice as large as between ranks 2 and 3. Therefore, it is likely that a larger sample size would have been required to observe a statistically significant difference between ranks 2 and 3.

\begin{tcolorbox}[title=Summary of Findings, colback=blue!3!white, colframe=blue!15!white,fonttitle=\bfseries, coltitle=black!85!white,boxsep=1mm, left=1mm, right=1mm, top=0.55mm, bottom=0.5mm,before skip=4pt,
	after skip=8pt]
	Our analysis shows that the rank annotations provided by our human judges are consistent with StackExchange’s quality indicators, both in terms of accepted answers and voting ranks. This supports the validity of the rankings collected in our study.
\end{tcolorbox}

\noindent\textbf{Experiment 2: Ranking of Generated Comments}\\
In this experiment, we assess the ranking of \textit{generated} comments compared to other comment types. In Table~\ref{table-significance-types}, we observe that generated comments receive significantly lower (better) ranks than accepted and standard human comments.

To further validate this finding, we tested whether the distribution of judge-assigned ranks for generated comments differs from a uniform distribution. 
If the ranks were uniformly distributed, it would suggest randomness or lack of meaningful differentiation by the judges. 
We applied Pearson’s chi-squared test \cite{Pearson1900} to test the null hypothesis that distributions of ranks for generated comments is uniform. 
This hypothesis was rejected (\emph{p-val\textless1e-4}), indicating that rankings are not random.

\vspace{0.5em}
\textbf{Discussion of Construct Validity.}
While these results suggest that the generated comments are meaningfully distinguished and often preferred by judges, the construct validity of this finding must be examined. 
Although we took measures to control for superficial indicators, such as controlling comment length, formatting patterns, and randomizing comment presentation, it is possible to make guesses about generated comments based on their polished grammar and structure. 

We instructed the judges to disregard surface-level style and focus solely on semantic quality when ranking comments. However, we cannot fully rule out the possibility that their rankings were biased, consciously or unconsciously, by preferences for more fluent or polished language, rather than substantive content. 
If this bias occurred, it may suggest that generated comments are preferred not due to superior semantic insight, but because of smoother expression.

\emph{Yet, even if this is the case, the results still carry important implications:} 
if humans tend to rate well-phrased but potentially semantically inferior comments higher than less polished human ones, this highlights a significant risk associated with the uncritical use of generative AI in code review processes.
It underscores the need for further research on the persuasive power of language models, especially in settings where clarity can mask a lack of substance.

For future work, we envision controlling these possible confounding factors more tightly by passing human-authored comments through an AI-based "polishing" step, ensuring uniform surface-level quality across all comment types.
We would then experimentally verify that judges cannot distinguish between generated and polished human comments.
Such normalized comments can then be reused in our evaluation framework, allowing for a purer focus on semantic quality.

\begin{tcolorbox}[title=Summary of Findings, colback=blue!3!white, colframe=blue!15!white,fonttitle=\bfseries, coltitle=black!85!white,boxsep=1mm, left=1mm, right=1mm, top=0.55mm, bottom=0.5mm,before skip=4pt,
	after skip=8pt]
	Based on our data, we conclude that ChatGPT’s generated comments are relevant with respect to a human standard, as captured by our construct. 
	However, we acknowledge potential confounding factors related to writing style and structure, for which we suggest a mitigation strategy to use in future work.
\end{tcolorbox}
\section{Threats to Validity}
Apart from the question of construct validity, discussed in detail in Section \ref{sec-evaluation}, we identify these additional threats.

\textbf{Internal validity.} 
We used sound statistical methodology and reported only strong signals as significant. 
Still, our results may be affected by unknown confounders or biases.

\textbf{External validity.}
Our study uses a medium-sized dataset and a small group of annotators, which may limit generalizability.
However, we observed significant alignment between our annotators and the platform's key quality signals.

Additionally, Stack Exchange reviews differ from code review practices on platforms such as GitHub or Bitbucket.

Finally, ChatGPT-3.5 Turbo has since been surpassed by more capable models. 
However, beyond the results for any specific model, our contribution is a general methodology that can be applied to any generative model.

\section{Conclusion}
In this paper, we introduced a new method to assess the relevance of AI-generated code reviews with respect to a human standard.
Compared to previous approaches, we use multi-subjective ranking to compare generated comments against multiple human-authored references, rather than relying on a single ground-truth comment.
To support this, we constructed an evaluation dataset of fully self-contained review requests from CodeReview StackExchange, thereby also mitigating the advantage human reviewers typically have from broader contextual knowledge.

Our statistical analyses demonstrated a significant alignment between annotator rankings and StackExchange's quality signals, and revealed that comments generated by ChatGPT were, on average, ranked significantly better than even the platform's accepted answers.
This suggests that, according to the employed construct, generative models are capable of producing relevant and useful review comments.

Finally, we discussed the threat that subjective preferences for superficial features, such as style and language, pose to construct validity.
Beyond proposing a mitigation strategy, this highlights a broader concern that generative language models may influence decisions based on surface-level fluency rather than actual content merit.

In conclusion, our work underscores the importance of empirical analyses for understanding the reliability of generative AI and ensuring its safe use in software engineering.

\textbf{Acknowledgments}
Removed for blind review.

\bibliographystyle{IEEEtran}
\bibliography{survey2025.bib, literature.bib}

\end{document}